\documentclass[11pt]{amsart}

\usepackage{latexsym}
\usepackage{amsfonts}
\usepackage{amsthm}
\usepackage{graphicx}
\usepackage{amssymb}
\usepackage{amsmath}
\usepackage{amssymb}
\usepackage{color}
\author[]{Dima Grigoriev}
\address{CNRS, Math\'ematiques, Universit\'e de Lille, 59655, Villeneuve d'Ascq, France}
\email{dmitry.grigoryev@math.univ-lille1.fr}
\thanks{Research of the first author was partially supported by the Federal Agency of the
Science and Innovations of Russia, State Contract No.
02.740.11.5192}

\author[]{Vladimir Shpilrain}
\address{Department of Mathematics, The City  College  of New York, New York,
NY 10031}
\email{shpil@groups.sci.ccny.cuny.edu}
\thanks{Research of the second author was partially supported by
the NSF grants DMS-0914778  and CNS-1117675}

\newtheorem{problem}{Problem}
\newtheorem{example}{Example}

\newtheorem{proposition}{Proposition}
\newtheorem{remark}{Remark}

\begin{document}

\title[Tropical cryptography]{Tropical cryptography}

\begin{abstract}
We employ tropical algebras as platforms for several cryptographic
schemes that would be vulnerable to linear algebra attacks were they
based on ``usual" algebras as platforms.

\end{abstract}

\maketitle

\medskip


\section{Introduction}

In this paper, we employ {\it tropical algebras} as platforms for
several cryptographic schemes. The  schemes themselves are not brand
new; similar ideas were used in the ``classical" case, i.e., for
algebras with the familiar addition and multiplication. However, in
the classical case these schemes were shown to be vulnerable to
various linear algebra attacks. Here we make a  case for using
tropical algebras as platforms by using, among other  things, the
fact that in the ``tropical" case, even solving systems of linear
equations is computationally infeasible in general. Yet another
advantage  is improved efficiency because in tropical schemes, one
does not have to perform any multiplications of numbers since
tropical multiplication is the usual addition, see below.

We start by giving some necessary information on tropical algebras
here; for more details, we refer the reader to a recent monograph
\cite{Butko}.

Consider a tropical semiring $S$ (also known as the min-plus algebra
due to the following definition). This semiring is defined as a
subset of reals that contains 0 and  is closed under addition, with
two operations as follows:
\medskip

\noindent $x \oplus y = \min(x, y)$

\medskip

\noindent $x \otimes y = x+y$.
\medskip

It is straightforward to see that these operations satisfy the
following properties:
\medskip

\noindent {\it associativity}: \\
$x \oplus (y  \oplus z) = (x \oplus y)  \oplus z$\\
$x \otimes (y  \otimes z) = (x \otimes y)  \otimes z$.

\medskip

\noindent {\it commutativity}:\\
$x \oplus y = y \oplus x$\\
$x \otimes y = y \otimes x$.
\medskip

\noindent {\it distributivity}:\\
$(x \oplus y)\otimes z = (x \otimes z)  \oplus (y\otimes z)$.

\medskip

There are some ``counterintuitive" properties as well:

\noindent $x \oplus x = x$
\medskip

\noindent $x \otimes 0 = x$
\medskip

\noindent $x  \oplus 0$ could be either 0 or $x$.

\medskip

There is also a special ``$\epsilon$-element"  $\epsilon = \infty$
such that, for any $x \in S$,
\medskip

\noindent $\epsilon \oplus x = x$
\medskip

\noindent $\epsilon \otimes x = \epsilon$.
\medskip

A  (tropical)  monomial in $S$ looks like a usual linear function,
and a tropical  polynomial is the minimum of a finite number of such
functions, and therefore a concave, piecewise linear function. The
rules for the order in which tropical operations are performed are
the same as in the classical case, see the example below.

\begin{example}
Here is an example of a   tropical monomial: $x \otimes x \otimes y
\otimes z \otimes z$. The  (tropical)  degree of this monomial is 5.
We note that sometimes, people use the alternative notation
$x^{\otimes 2}$ for $x \otimes x$, etc.

An example of a   tropical polynomial is: $p(x, y, z) = 5\otimes x
\otimes y  \otimes z  \oplus x \otimes x \oplus 2\otimes z \oplus 17
=  (5\otimes x \otimes y \otimes z) \oplus (x \otimes x) \oplus
(2\otimes z) \oplus 17.$ This polynomial has  (tropical) degree 3,
by the highest degree of its monomials.

We note that, just as in the classical case, a tropical polynomial
is canonically represented by an ordered set of tropical monomials
(together with non-zero coefficients), where the order that we use
here is  deglex.

\end{example}

While the  $\oplus$  operation is obviously not invertible, the
$\otimes$ operation is, and we denote the inverse of this operation
by $\oslash$ (it is just the classical subtraction):

\medskip

\noindent $x \oslash y = z$ if and only if $y \otimes z = x$.
\medskip

We refer to \cite{Mikhalkin} for more detailed properties of this
operation; here we just mention the following properties that agree
with those of the usual division:
\medskip

\noindent $(x \oslash y) \otimes (z \oslash t) = (x \otimes z)
\oslash (y \otimes t)$

\medskip

\noindent $(x \oslash y) \oplus (z \oslash t) = ((x \otimes t)
\oplus (y \otimes z)) \oslash (y \otimes t)$.
\medskip

Also as in the classical case, there is an  equivalence relation on
the set of all expressions of the form  $x \oslash y$:
\medskip

\noindent $x \oslash y$ is equivalent to $z \oslash t$ if and only
if ~$x \otimes t = y \otimes z$.
\medskip

All expressions of the form  $x \oslash y$, where $x, y \in S$,
modulo the above equivalence, form a semifield  (of quotients of
$S$), which we denote by $ Rat(S)$, see \cite{Mikhalkin}.


\subsection{Tropical matrix algebra}
\label{matrices}

A tropical algebra can be used for matrix operations as well. To
perform the $A \oplus B$  operation, the elements $m_{ij}$ of the
resulting matrix $M $   are set to be equal to $a_{ij} \oplus
b_{ij}$.  The  $\otimes$  operation is similar to the usual matrix
multiplication, however, every ``+" calculation has to be
substituted by a  $\oplus$ operation, and   every ``$\cdot$"
calculation by a  $\otimes$ operation.

\begin{example}
$\left(\begin{array}{cc} 1 & 2 \\ 5 & -1  \end{array}\right) \oplus
\left(\begin{array}{cc} 0 & 3 \\ 2 & 8  \end{array}\right) =
\left(\begin{array}{cc} 0 & 2 \\ 2 & -1  \end{array}\right).$

\end{example}

\begin{example}
$\left(\begin{array}{cc} 1 & 2 \\ 5 & -1  \end{array}\right) \otimes
\left(\begin{array}{cc} 0 & 3 \\ 2 & 8  \end{array}\right) =
\left(\begin{array}{cc} 1 & 4 \\ 1 & 7  \end{array}\right).$

\end{example}

The role of the identity matrix  $I$ is played by the matrix that
has ``0"s on the diagonal  and  $\infty$  elsewhere. Similarly, a
scalar matrix would be a matrix with an element $\lambda \in S$ on
the diagonal and  $\infty$  elsewhere. Such a matrix commutes with
any other square matrix (of the same size). Multiplying a square
matrix by a scalar amounts to multiplying it by the corresponding
scalar matrix.

\begin{example}

$2 \otimes \left(\begin{array}{cc} 1 & 2 \\ 5 & -1
\end{array}\right) = \left(\begin{array}{cc} 2 & \infty \\ \infty & 2
\end{array}\right)  \otimes  \left(\begin{array}{cc} 1 & 2 \\ 5 & -1
\end{array}\right) = \left(\begin{array}{cc} 3 & 4 \\ 7 & 1
\end{array}\right).$
\end{example}

Then, tropical {\it diagonal matrices} have something on the
diagonal and  $\infty$  elsewhere.

We also note that, in contrast with the ``classical" situation, it
is rather rare that a ``tropical" matrix is invertible. More
specifically (see \cite[p.5]{Butko}), the only invertible tropical
matrices are those that are obtained from  a diagonal matrix by
permuting  rows and/or columns.

\section{Key exchange using matrices over a tropical algebra}
\label{key_exchange}

We are now going to offer a key exchange protocol building on an
idea of Stickel \cite{Stickel} who used it for matrices over
``usual" algebras, which made his scheme vulnerable to linear
algebra attacks, see e.g. \cite{VS}. Since we believe that Stickel's
idea itself has a good potential, we suggest here to use matrices
over a tropical algebra as the platform for his scheme, in order to
prevent linear algebra attacks.

We start by recalling the original Stickel's protocol. Let $G$ be a
public non-commutative semigroup, $a, b \in G$ public elements such
that $ab \ne ba$. The key exchange protocol goes as follows.

\subsection{Protocol 1 \cite{Stickel}}

\begin{enumerate}

\item Alice picks two random natural numbers $n, m$ and sends $u=a^n
b^m$ to Bob.

\item    Bob picks two random natural numbers $r, s$ and sends
$v=a^r b^s$ to Alice.

\item   Alice computes $K_A=a^n v b^m = a^{n+r}b^{m+s}$.

\item  Bob computes $K_B=a^r u b^s = a^{n+r}b^{m+s}$.

\end{enumerate}

Thus, Alice and Bob end up with the same group element $K=K_A= K_B$
which can serve as the shared secret key.

This can be generalized if the platform is not just a semigroup, but
a ring  (actually, a semiring would suffice):

\subsection{Protocol 2 \cite{MMR, VS}}

Let $R$ be a public non-commutative ring (or a semiring), $a, b \in
R$ public elements such that $ab \ne ba$.

\begin{enumerate}

\item Alice picks two random polynomials $p_1(x), p_2(x)$ (say, with positive
integer coefficients) and sends $p_1(a) \cdot p_2(b)$ to Bob.

\item    Bob picks two random polynomials $q_1(x), q_2(x)$ and sends $q_1(a) \cdot q_2(b)$ to
Alice.

\item   Alice computes $K_A=p_1(a) \cdot (q_1(a) \cdot q_2(b)) \cdot
p_2(b)$.

\item  Bob computes $K_B=q_1(a) \cdot (p_1(a) \cdot p_2(b)) \cdot
q_2(b)$.

\end{enumerate}

Thus, since $p_1(a)\cdot q_1(a) = q_1(a) \cdot p_1(a)$ and
$p_2(b)\cdot q_2(b) = q_2(b) \cdot p_2(b)$,  Alice and Bob end up
with the same element $K=K_A= K_B$ which can serve as the shared
secret key.
\medskip

It is Protocol 2 that we propose to adopt in the ``tropical"
situation.

\subsection{Protocol 3 (tropical)}

Let $R$ be the tropical algebra of $n \times n$ matrices over
integers, and  let $A, B \in R$ be public matrices such that $A
\otimes  B \ne B  \otimes  A$.

\begin{enumerate}

\item Alice picks two random tropical  polynomials $p_1(x), p_2(x)$ (with
integer coefficients) and sends $p_1(A)  \otimes  p_2(B)$ to Bob.

\item    Bob picks two random tropical  polynomials $q_1(x), q_2(x)$ and sends
$q_1(A) \otimes  q_2(B)$ to
Alice.

\item   Alice computes $K_A=p_1(A)  \otimes  (q_1(A)  \otimes  q_2(B))  \otimes
p_2(B)$.

\item  Bob computes $K_B=q_1(A)  \otimes  (p_1(A) \otimes  p_2(B)) \otimes
q_2(B)$.

\end{enumerate}

Thus, since $p_1(A)\otimes q_1(A) = q_1(A) \otimes p_1(A)$ and
$p_2(B)\otimes q_2(B) = q_2(B) \otimes p_2(B)$,  Alice and Bob end
up with the same element $K=K_A= K_B$ which can serve as the shared
secret key.

\subsection{What are the advantages of the ``tropical" Protocol 3 over
``classical" Protocols  1 and  2?}

One obvious advantage is improved efficiency because when
multiplying matrices in the tropical sense, one does not have to
perform any multiplications of numbers since tropical multiplication
is the ``usual" addition.

To compare security, we briefly recall a linear algebra attack
\cite{VS} on Stickel's original protocol (Protocol 1), where $G$ was
a group of invertible matrices over a field. In that case, to
recover a shared key $K$, it is not necessary to find the exponents
$n, m, r$, or $s$. Instead, as was shown in \cite{VS}, it is
sufficient for the adversary to find matrices $x$ and  $y$ such that
$xa=ax, ~yb=by,$ and $xu=y$. (Here $x$ corresponds to $a^{-n}$,
while $y$ corresponds to $b^{m}$.)

  These
conditions translate into a system of $3k^2$ linear equations with
$2k^2$ unknowns, where $k$ is the size of the matrices. This
typically yields a unique solution (according to computer
experiments of \cite{VS} and \cite{Mullan}), which can be
efficiently found if the matrices are considered over a field.

We note that in \cite{Mullan}, a more sophisticated attack on a more
general Protocol 2 was offered. This attack applies to not
necessarily invertible matrices over a field.

In the ``tropical" situation (Protocol 3), however, a linear algebra
attack will not work, for several reasons:

\begin{enumerate}

\item Matrices are generically not invertible, so the equation $XY=U$ with known
$U$ and unknown $X, Y$ does not translate into a system of linear
equations.

\item The equations $XA=AX, ~YB=BY$ do translate into a system of linear
equations, which may be called a  ``two-sided min-linear system",
following \cite{Butko}. In \cite{Bezem}, it is shown that the
problem of solving such systems is  in the class ${\bf NP \cap
Co-NP}$ (there is a belief that it does not belong to the class $\bf
P$). We refer to \cite{Butko} for a comprehensive exposition of what
is known concerning existing algorithms for solving two-sided
min-linear  systems and their complexity. Here we just say that,
while it is known how to find one of the solutions of a system (if a
solution exists), there is no known efficient method for describing
the linear space of {\it all} solutions, in contrast with the
``classical" situation.


\end{enumerate}

\subsection{Parameters and key generation}
\label{parameters}

Here we suggest values of the parameters involved in the description
of our Protocol 3.

\begin{itemize}

\item The size of matrices $n=10$.

\item The entries of the public matrices $A, B$ are integers,
selected uniformly randomly in the range $[-10^{10}, 10^{10}]$.

\item The degrees of the tropical polynomials $p_1(x), p_2(x), q_1(x),
q_2(x)$ are selected  uniformly randomly in the range $[1, 10]$.

\item The coefficients of the above tropical polynomials  are selected  uniformly randomly in the range
$[-1000, 1000]$.

\end{itemize}

With these parameters, the size of the key space (for private
tropical polynomials) is approximately  $10^{30}$.

\section{Encryption using birational automorphisms of a tropical polynomial algebra}
\label{automorphisms}

In this section, we describe a public key encryption scheme that
would be susceptible to a linear algebra attack in the ``classical"
case (cf. \cite{Moh},  \cite{Goubin}), but not in the ``tropical"
case.

Let $P=Rat[x_1,\ldots, x_n]$ be the quotient semifield of a tropical
polynomial algebra over ${\mathbf Z}$.

\subsection{The protocol}

There is a public automorphism $\alpha \in Aut(P)$ given as a tuple
of tropical rational functions   $(\alpha(x_1),\ldots,
\alpha(x_n))$. Alice's private key is $\alpha^{-1}$. Note that
$\alpha$ is also a bijection of the set ${\mathbf Z}^n$, i.e., it is
a one-to-one map of the set of all $n$-tuples of integers onto
itself. We will use the same notation $\alpha$ for an automorphism
of $P$ and for the corresponding bijection of   ${\mathbf Z}^n$,
hoping this will not cause a confusion.

\begin{enumerate}

\item Bob's secret message is a tuple of integers  $s = (s_1,\ldots, s_n) \in {\mathbf
Z}^n$. Bob encrypts his  tuple by applying the public automorphism
$\alpha$: ~$E_\alpha(s)  = \alpha(s_1,\ldots, s_n)$.

\item Alice decrypts by applying her private $\alpha^{-1}$ to the tuple $E_\alpha(s)$:
$\alpha^{-1}(E_\alpha(s)) =s=(s_1,\ldots, s_n)$.

\end{enumerate}


\subsection{Key generation}
\label{parameters2}

The crucial ingredient in this scheme is, of course, generating the
public key $\alpha \in Aut(P)$. Alice can generate her automorphism
$\alpha$ as a product of ``monomial" automorphisms on the set of
variables $\{x_{1}, \ldots, x_n \}$ and  ``triangular" automorphisms
of the form

$$\varphi: x_i \to x_i \otimes  p_i(x_{i+1}, \ldots,  x_n), ~1 \le i \le n,$$

\noindent where $p_i \in P=Rat[x_1,\ldots, x_n]$.
Each   triangular automorphism, in turn, is a product of
``elementary" triangular automorphisms; these are of the form

$$\tau: x_j \to x_j \otimes  q_j(x_{j+1}, \ldots,  x_n), ~x_k  \to x_k, k \ne j.$$

\noindent The inverse of such a $\tau$ is

$$\tau^{-1}: x_j \to x_j  \oslash  (q_j(x_{j+1}, \ldots,  x_n)), ~x_k  \to x_k, k \ne j,$$

\noindent where $q_j \in P$.

``Monomial" automorphisms are analogs of linear automorphisms in the
``classical" situation; they are of the form

$$\mu: x_i \to b_i  \otimes x_1^{\otimes a_{i1}} \otimes \dots  \otimes x_n^{\otimes a_{in}},$$

\noindent where $b_i$ are finite coefficients (i.e., $b_i \ne
\infty$), and the matrix $A=(a_{ij})$ of integer exponents is
invertible in the ``classical" sense.

We note, in passing, that a question of independent interest
(independent of cryptographic applications) is:

\begin{problem}
Is every automorphism of $P=Rat[x_1,\ldots, x_n]$, the quotient
semifield of a tropical polynomial algebra over  ${\mathbf Z}$,  a
product of triangular and monomial  automorphisms?

\end{problem}



\subsection{Parameters}
\label{parameters3}

We suggest the following parameters.

\begin{itemize}

\item  The number $n$ of variables in the platform tropical polynomial
algebra: ~10.

\item  The number of triangular automorphisms in a product for
$\alpha$: ~2. The number of monomial automorphisms: ~3. More
specifically, Alice generates her $\alpha$ in the following form:
$$\alpha = \mu_1 \circ \varphi_1 \circ \mu_2 \circ \varphi_2 \circ \mu_3,$$

\noindent where $\varphi_1, \varphi_2$ are triangular automorphisms,
and $\mu_1, \mu_2, \mu_3$ are monomial automorphisms.

\item  The tropical degrees of all  $q_j$ are equal to 2.

\item  The coefficients of the above tropical polynomials $q_j$  are selected
uniformly randomly in the range $[-10, 10]$.

\end{itemize}

\begin{remark}
Alice can obtain the inverse of $\alpha$ as the product of inverses
of the  automorphisms $\varphi_i$ and  $\mu_i$, in the reverse
order. However, Alice does not have to compute an explicit
expression for $\alpha^{-1}$; this computation may not be efficient
since the degree of $\alpha^{-1}$ may be substantially greater than
the degree of $\alpha$. In our protocol, Alice has to apply
$\alpha^{-1}$ to a particular point in  ${\mathbf Z}^n$;  efficient
way of doing this is to first apply $\mu_3^{-1}$, then apply
$\varphi_2^{-1}$ to the obtained point, etc.

\end{remark}

\begin{remark}
There is a ramification of the above protocol, where Bob's secret
message is a tropical polynomial $u$, instead of a point in
${\mathbf Z}^n$. (Note that the result of encrypting $u$ will be, in
general, an element of $P=Rat[x_1,\ldots, x_n]$.) In this
ramification, decryption is going to have a much higher
computational complexity because Alice would have to compute an
explicit expression for $\alpha^{-1}$ (cf. the previous remark). On
the other hand, encryption in this case  is going to be {\it
homomorphic} (in the ``tropical" sense) because $\alpha(u_1 \oplus
u_2) = \alpha(u_1) \oplus \alpha(u_2)$ and $\alpha(u_1 \otimes u_2)
= \alpha(u_1) \otimes \alpha(u_2)$. For examples of homomorphic
encryption in the ``classical" case see e.g. \cite{Grigoriev2} or
\cite{Menezes}.

\end{remark}

\begin{remark}
One can consider an encryption protocol, similar to the one above,
also in the ``classical" case. As we have already pointed out,
polynomial automorphisms were employed in a similar context in
\cite{Moh}, but birational automorphisms have not been used for
cryptographic purposes before, to the best of our knowledge.

\end{remark}

\subsection{Possible attacks}
\label{attacks}

There are the following two attacks that adversary may attempt.

\begin{enumerate}

\item Trying to compute $\alpha^{-1}$ from the public automorphism
$\alpha$. The problem with this attack is that the degree of
$\alpha^{-1}$ may be exponentially greater than the degree of
$\alpha$, which makes any commonly used attack  (e.g. a linear
algebra attack) infeasible.

\item Trying to recover Bob's secret message   $s$ from
$\alpha(s)$. This translates into a system of tropical polynomial
equations; solving such a system is an NP-hard problem, as we show
in the following proposition.

\end{enumerate}

\begin{proposition}
The problem of solving systems of tropical polynomial equations is
NP-hard.
\end{proposition}

Before getting to the proof, we note that for a closely related, but
different, problem of emptiness of a tropical variety
NP-completeness was established in \cite{TT}.

\begin{proof}
We show how to reduce the SAT problem to the problem of solving a
system of tropical polynomial equations. Recall that the SAT (for
SATisfiability) problem is a decision problem, whose instance is a
Boolean expression written using only AND, OR, NOT, variables, and
parentheses. The question is: given the expression, is there some
assignment of TRUE (=1) and FALSE (=0) values to the variables that
will make the entire expression true? A formula of propositional
logic is said to be satisfiable if logical values can be assigned to
its variables in a way that makes the formula true. The Boolean
satisfiability problem is NP-complete \cite{GJ}. The problem remains
NP-complete even if all expressions are written in conjunctive
normal form with 3 variables per clause (3-CNF), yielding the 3-SAT
problem.

Suppose now we have a 3-CNF, and we are going to build (in time
polynomial in the number of clauses) a system of tropical polynomial
equations that has a solution if and only if the given 3-CNF is
satisfiable. Denote Boolean variables in the given 3-CNF by $u_i$.
In our tropical system, we are going to have two kinds of variables:
those corresponding to literals $u_i$ will be denoted by $x_i$, and
those corresponding to literals $\neg u_i$ will be denoted by $y_i$.

First of all, we include in our tropical system all equations of the
form ~$x_i  \otimes y_i = 1$, ~for all $i$.

Now suppose we have a clause with 3 literals, for example,  $u_i
\vee \neg u_j \vee \neg u_k$. To this clause, we correspond the
following  tropical polynomial equation:

$$y_i \oplus x_j \oplus x_k = 0.$$

\noindent Obviously, the above clause is TRUE if and only if either
$u_i=1$, or  ~$u_j=0$,   or  ~$u_k=0$. If  $u_i=1$, then $y_i=0$,
and our tropical equation is satisfied. If, say,  $u_j=1$, then
$x_j=0$, and again our tropical equation is satisfied. This shows
that if a given 3-CNF is satisfiable, then our tropical equation has
a solution.

If, on the other hand,  our tropical equation has a solution, that
means either $y_i=0$, or  ~$x_j=0$,   or  ~$x_k=0$. In any case, the
given clause is easily seen to be TRUE upon corresponding $u_i$ to
$x_i$ and $\neg u_i$ to $y_i$. (Note that if, say, $y_i=0$, then,
since we also have the equation $x_i  \otimes y_i = 1$, ~$x_i$
should be equal to 1.)


Having thus built a tropical equation for each clause in the given
3-CNF, we end up with a system of tropical polynomial equations that
corresponds to the whole 3-CNF, which is solvable if and only if the
given 3-CNF is satisfiable.
 This completes the proof.

\end{proof}

\vskip .5cm

\noindent {\it Acknowledgement.} Both authors are grateful to  Max
Planck Institut f\"ur Mathematik, Bonn for its hospitality during
the work on this paper.


\baselineskip 11 pt


\begin{thebibliography}{ABC}


\bibitem{Bezem}
M. Bezem, R. Nieuwenhuis,  E. Rodríguez-Carbonell,  {\it  Hard
problems in maxalgebra, control theory, hypergraphs and other
areas}, Information Processing Letters {\bf  110(4)} (2010),
133–-138.

\bibitem{Butko}
P. Butkovic,  {\em  Max-linear systems: theory and algorithms},
Springer-Verlag London, 2010.


\bibitem{GJ}
M. Garey, J. Johnson, {\em Computers and Intractability, A Guide to
NP-Completeness}, W. H. Freeman, 1979.

\bibitem{Goubin}
L. Goubin, N. Courtois,  {\it Cryptanalysis of the TTM
cryptosystem}, in: ASIACRYPT 2000, Lecture Notes in Comput. Sci.
{\bf 1976} (2000),  44–-57.

\bibitem{Grigoriev2}
D. Grigoriev, I. Ponomarenko, {\it Constructions in public-key
cryptography over matrix groups},  Contemp. Math., Amer. Math. Soc.
{\bf  418} (2006), 103--119.

\bibitem{MMR}
G. Maze, C. Monico, J. Rosenthal, {\it Public key cryptography based
on semigroup actions}, Advances in Mathematics of Communications
{\bf  4} (2007), 489–-507.

\bibitem{Menezes}
A. Menezes, P. van Oorschot, and S. Vanstone, {\sl  Handbook of
Applied Cryptography}, CRC-Press 1996.

\bibitem{Mikhalkin}
G. Mikhalkin, {\sl Tropical geometry}, in preparation.\\
http://www.math.toronto.edu/mikha/book.pdf


\bibitem{Moh}
T. Moh,  {\it  A public key system with signature and master key
functions},  Comm. Algebra {\bf  27} (1999),  2207–-2222.

\bibitem{Mullan}
C. Mullan, {\it Cryptanalysing variants of Stickel's key agreement
scheme}, preprint.


\bibitem{VS}
V. Shpilrain, {\it Cryptanalysis of Stickel's key exchange scheme},
in: Computer Science in Russia 2008, Lecture Notes Comp. Sc. {\bf
5010} (2008), 283–-288.

\bibitem{SSC}
R. Steinwandt and A. Su\'arez Corona, {\it  Cryptanalysis of a
2-party key establishment based on a semigroup action problem},
Advances in Mathematics of Communications  {\bf 5}  (2011), 87--92.

\bibitem{Stickel}
E. Stickel, {\it  A New Method for Exchanging Secret Keys.} In:
Proc. of the Third International Conference on Information
Technology and Applications (ICITA'05) 2 (2005), 426--430.

\bibitem{TT}
T. Theobald,  {\it   On the frontiers of polynomial computations in
tropical geometry}, J. Symbolic Comput. {\bf 41} (2006), 1360--1375.


\end{thebibliography}
\end{document}